\documentclass{aastex62}

\graphicspath{{./}{figures/}}

\received{August 28, 2018}
\revised{August 30, 2019}
\accepted{August 30, 2019}
\submitjournal{ApJ}

\shorttitle{The reactive relativistic Riemann problem}
\shortauthors{Harpole \& Hawke}

\usepackage{mathtools}
\usepackage{savesym}
\savesymbol{tablenum}
\usepackage{siunitx}
\restoresymbol{SIX}{tablenum}
\DeclareSIUnit[number-unit-product = {}]\erg{erg}
\DeclareSIUnit[number-unit-product = {}]\year{yr}
\DeclareSIUnit[number-unit-product = {}]\gauss{G}
\DeclareSIUnit[number-unit-product = \,]\solarM{\ensuremath{\mathrm{M}_\odot}}

\usepackage{tikz}
\usepackage{pgfgantt}
\usepackage{pgfplots}
\pgfplotsset{compat=1.14}
\usetikzlibrary{external,arrows.meta,arrows,patterns,decorations,decorations.pathmorphing,decorations.text,decorations.shapes}
\usetikzlibrary{math,intersections}
\usepackage{xfrac}
\usepackage{stmaryrd}

\providecommand*{\bs}[1]{\boldsymbol{#1}}
\newcommand{\fref}[1]{Figure~\ref{#1}}
\newcommand{\sref}[1]{Section~\ref{#1}}
\newcommand{\tref}[1]{Table~\ref{#1}}

\usepackage[caption=false]{subfig}

\providecommand*{\partder}[3][]{\frac{\partial^{#1}#2}{\partial {#3}^{#1}}} 

\begin{document}

\title{Effects of tangential velocity in the reactive relativistic Riemann problem}

\correspondingauthor{A.~Harpole}
\email{alice.harpole@stonybrook.edu}

\author{A.~Harpole}
\affil{Mathematical Sciences and STAG Research Centre, \\
University of Southampton, \\
Southampton SO17 1BJ, UK}
\affiliation{Department of Physics and Astronomy, \\
Stony Brook University, \\
Stony Brook, NY 11794-3800 USA}

\author{I.~Hawke}
\affiliation{Mathematical Sciences and STAG Research Centre, \\
University of Southampton, \\
Southampton SO17 1BJ, UK}



\begin{abstract}
Type I X-ray bursts are thermonuclear burning events which occur on the surfaces of accreting neutron stars. Burning begins in a localised spot in the star's ocean layer before propagating across the entire surface as a deflagration. On the scale of the entire star, the burning front can be thought of as discontinuity. To model this, we investigated the reactive Riemann problem for relativistic deflagrations and detonations and developed a numerical solver. Unlike for the Newtonian Riemann problem, where only the velocity perpendicular to the interface is relevant, in the relativistic case the tangential velocity becomes coupled through the Lorentz factor and can alter the waves present in the solution. We investigated whether a fast tangential velocity may be able to cause a deflagration wave to transition to a detonation. We found that such a transition is possible, but only for tangential velocities that are a significant fraction of the speed of light or for systems already on the verge of transitioning. Consequently, it is highly unlikely that this transition would occur for a burning front in a neutron star ocean without significant contributions from additional multidimensional effects.
\end{abstract}

\keywords{stars: neutron -- X-rays: bursts -- stars: oscillations -- X-rays: binaries}

\section{Introduction}

Type I X-ray bursts are thermonuclear burning events which occur on the surfaces of accreting neutron stars (see \citet{Strohmayer2001,Cumming2004,IntZand2010} for reviews). Burning begins in a localised spot in the star's ocean layer before spreading across the entire surface \citep{Joss1978,Shara1982}. We observe the burning as a sharp increase in the X-ray radiation of the star. These bursts typically occur every few hours to days and release \(\sim10^{39} - \SI{e40}{\erg}\) energy. By gaining a better understanding of X-ray bursts, tighter limits can be determined for other neutron star properties such as the mass, radius, spin frequency and magnetic field \citep{Miller2013,Watts2016,Ozel2016}. 

The physics of X-ray bursts acts over a wide range of scales, which introduces a number of challenges when modelling them. During an X-ray burst, the burning front propagates as a deflagration through the ocean. Whilst we expect the motion of the front to be driven by turbulent combustion \citep{Reinecke1999}, the burning itself will be confined to a reaction zone no more than a few centimetres thick \citep{Hillebrandt2013}. When the front is viewed on length scales of the order the neutron star radius, the front can be treated as a discontinuity, with the burning reactions happening on very short timescales compared to the propagation of the front. We can consider the front to be sharp (i.e.~a discontinuous change in the matter fields) and governed by the equations of ideal (relativistic) hydrodynamics with a reaction term. In this case, we can approximate the system as a reactive Riemann problem.

The reactive Riemann problem was first considered by \citet{Zhang1989}, where they looked at the problem for gasdynamic combustion. They modelled the combustion process as the release of binding energy \(q\) through the burning of the reactive gas. They treat this process as if it has an infinite rate of reaction so that it happens instantaneously once the unburnt gas reaches a certain ignition temperature; after burning the temperature of the burnt gas may not remain higher than this. The solution to the Riemann problem with reactions may now include detonations and deflagrations, the reactive counterparts of shock and rarefactions with the additional release of binding energy. This shall be discussed in more detail in \sref{sec:reactive_rp}.

As found by \citet{Pons2000,Rezzolla2002}, special relativistic effects exist in the solution of the Riemann problem for inert relativistic hydrodynamics for which there are no Newtonian counterparts. Specifically, the qualitative development of the Riemann problem can depend on the magnitude of the component of the fluid velocity \emph{tangential} to the initial discontinuity. In the Newtonian case there are quantitative changes to, for example, shock speeds with changes to the tangential velocities (see, for example, the \emph{oblique shocks} discussed in Section 92 of~\cite{Landau1987}). However, a \emph{qualitative} change in the wave structure only occurs in relativity.
Such a jump in the tangential velocity could appear automatically at a burning front as it progresses across a rotating neutron star, as the reacted products behind the front will lead to small changes in the local mass, without changing the specific angular momentum due to the global rotation. We will show that changes from subsonic deflagrations to supersonic detonations can occur in relativity from the introduction of a non-zero tangential velocity, and discuss whether this may be relevant for the rapid fluctuations in the burning front associated with X-ray bursts.

We shall begin in \sref{sec:rp} by describing the Riemann problem, and its extension to the reactive Riemann problem in \sref{sec:reactive_rp} and the relativistic reactive Riemann problem in \sref{sec:rrrp}. In \sref{sec:numerics}, we shall present results produced by a numerical solver for the relativistic reactive Riemann problem, \texttt{R3D2}. A detailed description of this code was previously presented in \citet{Harpole2016}.
We will show that the introduction of tangential velocities is sufficient to change the wave pattern from deflagration to detonation and consider whether the magnitude of tangential velocity required for this transition could be achieved in neutron star oceans.

\section{The Riemann problem}\label{sec:rp}
The Riemann problem is defined as a system of hyperbolic conservation laws whose initial data $\bs{q}(0,x)$ consists of two constant states, $\bs{q}_L$, $\bs{q}_R$, separated by a single jump discontinuity \citep{Leveque2002},
\begin{equation}
\bs{q}(0,x) = \begin{cases} \bs{q}_L & \textrm{if } x < 0, \\
\bs{q}_R & \textrm{if } x > 0.\end{cases}
\end{equation}
The solution of the Riemann problem involves computing the breakup of the initial discontinuity. This consists of a set of waves separating a set of constant states. These waves can be \emph{simple waves}, which can take the form of nonlinear waves (such as shocks or rarefactions) and linear waves (such as contact discontinuities and rotational linear waves in MHD). As shall be seen for the reactive Riemann problem, the solution may also contain \emph{compound waves} if the flux is non-convex. These consist of multiple simple waves `attached' to each other, moving as one.  The exact nature of the wave pattern, including the number of waves in the final state, depends on the initial conditions of the problem. The solution is self-similar, such that it can be written as a function of a single independent variable,
\begin{equation}\label{eq:xi}
\xi = \frac{x}{t}.
\end{equation}
Waves in the solution follow lines of constant $\xi$. In the case of a non-reactive fluid, the discontinuity will decay into two nonlinear waves moving in opposite directions with respect to the fluid flow. Between these two waves will be two new constant states $\bs{q}^*_{L}$, $\bs{q}^*_{R}$, separated by a contact discontinuity that moves with the fluid velocity.

We wish to solve the Riemann problem for the non-relativistic Euler equations
\begin{equation}\label{eq:euler}
\partder{\bs{q}}{t} + \partder{\bs{f}^{(i)}(\bs{q})}{x^i} = 0,
\end{equation}
where 
\begin{equation}
\bs{q} = \left(\rho, \rho v^j, E\right)^\intercal
\end{equation}
is the state vector of conserved variables. \(\rho\) is the density, \(v^j\) the fluid velocity in the \(j\)-direction and \(E = \rho e + \rho v^2/2\) is the total energy, where \(e\) is the specific internal energy.  The corresponding vector of fluxes is then 
\begin{equation}
\bs{f}^{(i)}\left(\bs{q}\right) = \left(\rho v^i, v^iv^j + p \delta^{ij}, (E+p)v^i\right)^\intercal.
\end{equation}
The system is closed by an equation of state, \(e = e(p, \rho)\).

\begin{figure} \centering 
\begin{tikzpicture}[scale=0.8]

	\tikzset{
	every pin/.style={font=\small, pin distance=2mm},
	small dot/.style={fill=black,circle,scale=0.3}
	}

	\draw[-latex] (-5,0) -- (5,0) node[right] {$x$};
	\draw[-latex] node[below, font=\Large] {$I$} (0,0) -- (0,5) node[above] {$t$};
	
	\draw[thick, color=blue] (0,0) -- (4,4) node[right, color=black, font=\Large] {$\mathcal{W}_\rightarrow$};
	\draw[thick, color=blue] (0,0) -- (-3,4) node[left, color=black, font=\Large] {$\mathcal{W}_\leftarrow$};
	\draw[thick, color=blue, dashed] (0,0) -- (1.5,4.5) node[above, color=black, font=\Large] {$\mathcal{C}$};
	
	\node[font=\Large] at (-3, 1.5) {$L$};
	\node[font=\Large] at (3.5,1.5) {$R$};
	\node[font=\Large] at (-1, 3.5) {$L_*$};
	\node[font=\Large] at (2, 3.3) {$R_*$};
\end{tikzpicture}

\caption{Wave pattern for the Riemann problem of a non-reactive fluid with initial left and right states $L$, $R$. The initial discontinuity decays into two nonlinear waves $\mathcal{W}_\leftarrow$, $\mathcal{W}_\rightarrow$, between which there are two new constant states $L_*$, $R_*$ separated by a linear contact discontinuity $\mathcal{C}$.}
\label{fig:riemannproblem}
\end{figure}
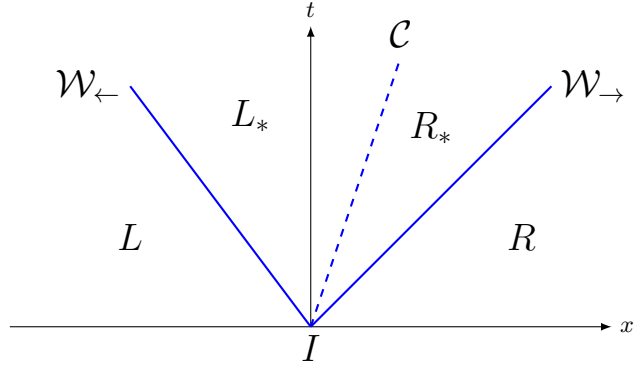

We can represent the time evolution of the Riemann problem in a non-reactive fluid with the initial state $I$ as
\begin{equation}
\label{eq:rp_state_evolution}
I \rightarrow L \mathcal{W}_\leftarrow L_* \mathcal{C} R_* \mathcal{W}_\rightarrow R,
\end{equation}
where $\mathcal{W}$ denotes a nonlinear simple wave (a shock or a rarefaction) and $\mathcal{C}$ a contact discontinuity. The states are labelled from left to right by $L, L_*, R_*$ and $R$.  The arrows represent the direction from which a fluid element enters the wave. This solution is illustrated in \fref{fig:riemannproblem}.

The distinction between shock and rarefaction waves is determined by the difference in pressure between the left and right states, such that
\begin{equation}
\mathcal{W} = \begin{cases}\mathcal{R}, & p_b \leq p_a,\\
\mathcal{S}, & p_b > p_a.
\end{cases}
\end{equation}
The subscripts indicate quantities ahead of and behind the wave: $a \equiv L(R)$ and $b \equiv L_*(R_*)$ for $\mathcal{W}_\leftarrow (\mathcal{W}_\rightarrow)$. The initial discontinuity can decay in three distinct ways,
\begin{equation}
I \rightarrow \begin{cases} L \mathcal{S}_\leftarrow L_* \mathcal{C}R_*\mathcal{S}_\rightarrow R, \quad &\text{if } p_L < p_{*} \text{ and } p_R < p_{*},\\
L \mathcal{S}_\leftarrow L_* \mathcal{C}R_*\mathcal{R}_\rightarrow R, &\text{if } p_L < p_{*}\leq p_R,\\
 L \mathcal{R}_\leftarrow L_* \mathcal{C}R_*\mathcal{R}_\rightarrow R, &\text{if } p_* \leq p_L \text{ and } p_{*} \leq p_R,
\end{cases}
\end{equation}
where $p_{L*} = p_{R*} = p_*$. Which of these three wave patterns the solution takes depends on the initial data.

\subsection{Exact solution}

For the inert non-relativistic Riemann problem, it is possible to find an exact solution if the initial states $\bs{q}_L$ and $\bs{q}_R$ are known. As the pressure and normal velocity are constant across the contact wave, we can find the solution by solving an implicit equation for the pressure in the star states $p_*$ \citep{Toro1999},
\begin{equation}
f(p_*, \bs{q}_L, \bs{q}_R) \equiv f_L(p_*, \bs{q}_L) + f_R(p_*, \bs{q}_R) + \Delta v = 0,
\end{equation}
where $\Delta v \equiv v_{x,R} - v_{x,L}$ is the difference across the wave of the normal velocity, $v_x$. Denoting the left and right states $L$, $R$ by $S$, the functions $f_L$, $f_R$ are given by
\begin{equation}\label{eq:exact_shock}
f_S(p_*, \bs{q}_S) = \begin{cases} (p_* - p_S)\left(\frac{A_S}{p_* + B_S}\right)^{\frac{1}{2}} & \;\text{if } p_* > p_S \text{ (shock)},\\
\frac{2 c_{s,S}}{\gamma_S-1}\left[\left(\frac{p_*}{p_S}\right)^{\frac{\gamma_S-1}{2\gamma_S}} - 1\right] &\; \text{if } p_* \leq p_S \text{ (rarefaction)},
\end{cases}
\end{equation}
where the constants $A_S$, $B_S$ are given by
\begin{equation}
A_S = \frac{2}{(\gamma_S+1)\rho_S},\qquad B_S = \frac{\gamma_S-1}{\gamma_S+1}p_S.
\end{equation}
The adiabatic index $\gamma_S = c_{p,S} / c_{V,S}$ is the ratio of specific heat capacities. It need not be the same in the left and right states -- this solution holds even if the initial states have different equations of state.  Once the pressure in the star states has been obtained, the other variables in the star states can then be calculated using the equation of state. 
The normal velocity in the star state $v_{x,*}$ is given by
\begin{equation}
v_{x,*} = \frac{1}{2}\left(v_{x,L} + v_{x,R}\right) + \frac{1}{2}\left[f_R\left(p_*\right) - f_L\left(p_*\right)\right].
\end{equation}

\section{The reactive Riemann problem}\label{sec:reactive_rp}

The nature of the thermonuclear burning that occurs during type I X-ray bursts is largely dependent on the composition of the accreted material and the accretion rate. The simplest case is that of a star accreting pure \({}^4\)He, where the burning proceeds via the \(3\alpha\) process; in bursts involving H, slower \(\beta\)-decay limited processes must also be taken into account \citep{Malone2011}. A simple model of \({}^4\)He burning is to use a two species system (rather than using a complex reaction network), consisting of the unburnt \({}^4\)He and the \({}^{12}\)C ashes \citep{Cumming2000c}. 
As described by \citet{Spitkovsky2002}, such a two species model neglects further energy release due to nuclear evolution beyond carbon, however it is sufficient for the purposes of our work. 

To model reaction terms, we can add a source term to the Euler equations \eqref{eq:euler} and extend the state vector to model the evolution of the baryon fraction \(X\) (assuming a two-species system where \(0 \leq X \leq 1\)). The Euler equations therefore become
\begin{equation}
\partder{\bs{q}}{t} + \partder{\bs{f}^{(i)}(\bs{q})}{x^i} = \bs{S},
\end{equation}
where the state vector of conserved variables is now 
\begin{equation}
\bs{q} = \left(\rho, \rho v^j, E, \rho X\right)^\intercal,
\end{equation}
the fluxes are given by 
\begin{equation}
\bs{f}^{(i)}(\bs{q}) = \left(\rho v^i, v^iv^j + p \delta^{ij}, (E+p)v^i, \rho X v^i\right)^\intercal
\end{equation}
and the source term is 
\begin{equation}
\bs{S} = \left(0,0,\rho q\dot{\omega}, \rho\dot{\omega}\right)^\intercal.
\end{equation}
Here, \(\dot{\omega}\) is the species creation rate of species \(X\) and \(q\) is the specific binding energy.

In the reactive Riemann problem, we are interested in systems where the reactions happen `instantly', i.e.~where the species creation rate $\dot{\omega} \to \infty$. If we work in the frame of the flow, we see that the jump in the energy is given by
\begin{equation}
  \label{eq:energy_jump}
  Q = \int_{X=0}^{X=1} \frac{\text{d}X}{\rho} = q,
\end{equation}
where $X$ is the species mass fraction and $q$ is the specific binding energy. Therefore, we can model the system by neglecting the source terms completely (as they model the reactions which are happening `instantly') and instead consider the equation of state to change across the sharp, nonlinear wave that models the reaction. Specifically, the total internal energy $e$ that specifies the equation of state will change to $e \to e - q$ when the reaction takes place. This is precisely the model considered by \citet{Zhang1989} when constructing their Newtonian reactive Riemann problem solution. As in their model, for simplicity we shall also make the assumption that the adiabatic exponents of both burnt and unburnt gases are the same.

\subsection{Detonations}
\label{sec:newt_detonation}

A detonation is a discontinuous reactive wave across which the pressure increases. The equations to be solved to find the change in variables across the wave are identical to the equations \eqref{eq:exact_shock} for shocks, but the interpretation changes. All `known' (pre-shock) variables have the reactive equation of state. All `unknown' (post-shock) variables use the inert equation of state -- the reaction has taken place across the discontinuity.

\begin{figure} \centering 
\subfloat[Strong detonation]{ \centering 
			\begin{tikzpicture}[scale=0.6]
			
				\clip (-5,-1) rectangle (6,6);
			
				\draw[-latex] (-5,0) -- (5,0) node[right] {$x$} ;
				\draw[-latex] (0,0) node[below] {0} -- (0,5) node[above] {$t$};
				
				\foreach \x in {1,2,3} {
					\path[name path=d] (-\x, 0) -- (4/0.5-\x, 4);
					\path[name path=e] (\x, 0) -- (4/3.0+\x, 4);
				
					\path [name intersections={of=d and e, by={a}}];
			
					\draw[blue] (-\x, 0) -- (a);
					\draw[blue] (\x,0) -- (a);
				}
				
				\path[name path=d] (-4, 0) -- (4.2/0.5-3.7, 4.2);
				\path[name path=e] (4, 0) -- (4.2/3.0+3.7, 4.2);
				\path [name intersections={of=d and e, by={a}}];
				\draw[very thick,red] (0, 0) -- (a) node[above, black] {$\mathcal{SDT}$};
			
				\node[font=\Large] at (-3, 1.5) {$\bs{q}_L$};
				\node[font=\Large] at (4.5, 1.5) {$\bs{q}_R$};
			\end{tikzpicture}
			\label{fig:strong_det}
	}
	\subfloat[CJ detonation + rarefaction]{ \centering 
			\begin{tikzpicture}[scale=0.6]
			
				\clip (-6,-1) rectangle (7.2,6);
			
				\draw[-latex] (-5,0) -- (5,0) node[right] {$x$} ;
				\draw[-latex] (0,0) node[below] {0} -- (0,5) node[above] {$t$};
				
				\foreach \x in {1,2,3,4} {
					\path[name path=d] (0, 0) -- (8,8*0.7);
					\path[name path=e] (\x, 0) -- (6, 6*4-\x*4);
				
					\path [name intersections={of=d and e, by={a}}];
			
					\draw[blue] (\x,0) -- (a);
				}
				
				\foreach \x in {1,2,3} {
					\draw[blue] (-\x-1, 0) -- (4.2/2-\x-1, 4.2);
				}
				
				\draw[olive!30!green!80!white,very thick,dotted] (-0.6*0.7,0) -- (5.3-0.6*0.7, 5.3*0.7);
				
				\node at (2.5, 4.5) {$\mathcal{R}$};

				\draw[very thick,red] (0, 0) -- (a) node[right, black] {$\mathcal{CJDT}$};
				
				\draw[very thick] (-1, 0) -- (4.1/0.7-1,4.1);
				
				\draw[very thick] (-1, 0) -- (4.2/2-1, 4.2);
				
				\draw[blue, dashed] (-1,0) -- (4.2-1, 4.2*0.7);
				\draw[blue, dashed] (-1,0) -- (4.2/1-1, 4.2);
				\draw[blue, dashed] (-1,0) -- (4.2/1.5-1, 4.2);
			
				\node[font=\Large] at (-4, 1.5) {$\bs{q}_L$};
				\node[font=\Large] at (5.5, 1.5) {$\bs{q}_R$};
				
			\end{tikzpicture}
			\label{fig:weak_det}
	}

\caption{Characteristics of detonations. For the strong detonation, characteristics (blue lines) from the initial left and right states meet, forming a strong detonation (thick red line). 
For the unstable weak detonation, the characteristics impinge only on the right side of the discontinuity. It is replaced by a compound wave made up of a CJ detonation (red line) and a rarefaction (thick black lines and dashed blue lines). The dotted green line between them illustrates that the characteristics to the left of the CJ detonation are parallel to the discontinuity -- in reality, the width of this region shrinks to zero and the detonation and rarefaction waves are attached together.}
\label{fig:detonation_characteristics}
\end{figure}
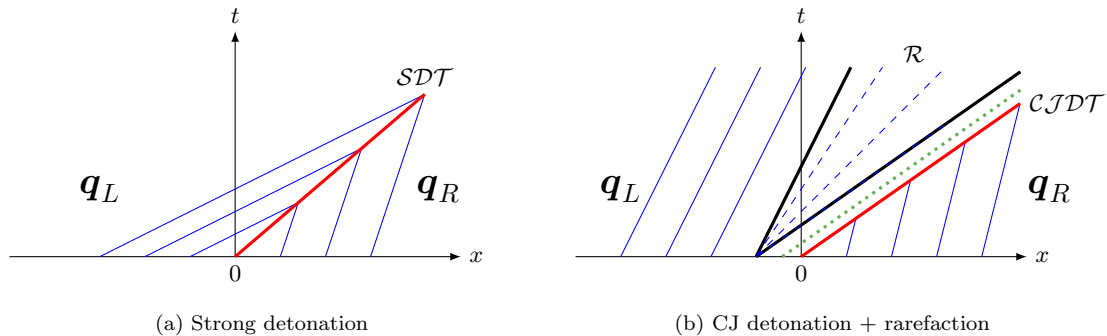

It is possible that the resulting detonation wave is unstable. In general, detonations fall into two classes: unstable weak detonations and stable strong detonations. For the stable strong detonations, the characteristic waves impinge on the discontinuity from both sides (see \fref{fig:strong_det}). They are \emph{compressive} and, in the frame of the discontinuity, supersonic. For the unstable weak detonations, the characteristics only enter the discontinuity on one side.

If the solution takes the form of an unstable weak detonation, the single detonation wave is replaced by a compound wave. The compound wave is composed of the fastest detonation wave that is stable, a Chapman-Jouget (CJ) detonation, and a rarefaction. This CJ detonation is where the characteristic waves are parallel to the discontinuity, propagating at the same speed. It can be thought of the `weakest' possible strong detonation. As the post-detonation pressure will now no longer match the required post-wave pressure, the additional rarefaction wave is needed. This is illustrated in \fref{fig:weak_det}.

Using the calligraphic notation used above to represent the time evolution of the Riemann problem, we shall denote strong detonations by $\mathcal{SDT}$ and CJ detonations by $\mathcal{CJDT}$.

\subsection{Deflagrations}
\label{sec:newt_deflagration}

A deflagration is a discontinuous reactive wave across which the pressure decreases. If the equation of state is convex and the reaction exothermic, then across a reactive discontinuous wave the pressure increases. Consequently, reactions cannot happen across a rarefaction wave and a discontinuity is required. However, this discontinuity will reduce the temperature along with the pressure. This means that, unless the material was already at the right temperature to react, any reaction across this wave would be unphysical. The solution for a deflagration therefore requires a compound wave.

\begin{figure} \centering 
	\subfloat[Weak deflagration] {\centering 
			\begin{tikzpicture}[scale=0.6]
				
				\clip (-5,-1) rectangle (7,6);
			
				\draw[-latex] (-4,0) -- (5,0) node[right] {$x$} ;
				\draw[-latex] (0,0) node[below] {0} -- (0,5) node[above] {$t$};
				
				\foreach \x in {1,2,3} {
					\path[name path=d] (-\x, 0) -- (4/0.5-\x, 6);
					\path[name path=e] (\x, 0) -- (4/6.0+\x, 6);
				
					\path [name intersections={of=d and e, by={a}}];
			
					\draw[blue] (-\x, 0) -- (a);
					\draw[blue] (\x,0) -- (5, 5*0.6-\x*0.6);
				}
				
				\path[name path=d] (-4, 0) -- (4.2/0.5-3.7, 4.2);
				\path[name path=e] (4, 0) -- (4.2/3.0+3.7, 4.2);
				\path [name intersections={of=d and e, by={a}}];
				\draw[very thick,red] (0, 0) -- (a);
				\node at (6, 4.2) {$\mathcal{WDF}$};
				
				\draw[very thick,red] (0, 0) -- (5, 5*0.6);
				
				\draw[red, dashed] (0,0) -- (5, 5*0.8);
				\draw[red, dashed] (0,0) -- (4.6/1, 4.6);
				\draw[red, dashed] (0,0) -- (4.8/1.2, 4.8);
			
				\node[font=\Large] at (-3, 1.5) {$\bs{q}_L$};
				\node[font=\Large] at (6, 1.5) {$\bs{q}_R$};
			\end{tikzpicture}
			\label{fig:weak_def}
}
\subfloat[CJ deflagration + rarefaction] { \centering 
		\tikzsetnextfilename{strong_def_characteristics}
			\begin{tikzpicture}[scale=0.6]
			
			\clip (-5,-1) rectangle (6.5,6);
			
				\draw[-latex] (-5,0) -- (5,0) node[right] {$x$} ;
				\draw[-latex] (0,0) node[below] {0} -- (0,5) node[above] {$t$};
				
				\foreach \x in {1,2,3} {
					
					\draw[blue] (\x,0) -- (4.5, -0.5*\x+4.5*0.5);
					\draw[blue] (-\x-1, 0) -- (4.2/3-\x-1, 4.2);
				}
				
				\draw[olive!30!green!80!white,very thick,dotted] (-0.6*0.7, 0) -- (4.2/1.2-0.6*0.7, 4.2);
				
				\draw[very thick] (-1, 0) -- (4.2/3-1,4.2);
				\draw[very thick] (-1, 0) -- (4.2/1.2-1,4.2);
				\draw[very thick,red] (0, 0) -- (4.5,4.5*0.5);
				\draw[very thick,red] (0, 0) -- (4.2/1.2,4.2);
				
				\draw[red, dashed] (0,0) -- (4.2, 4.2);
				\draw[red, dashed] (0,0) -- (4.5, 4.5*0.7);
				
				\draw[blue, dashed] (-1,0) -- (4.2/1.5-1, 4.2);
				\draw[blue, dashed] (-1,0) -- (4.2/2-1, 4.2);
			
				\node[font=\Large] at (-4.2, 1.5) {$\bs{q}_L$};
				\node[font=\Large] at (5.2, 1.5) {$\bs{q}_R$};
				
				\node at (5.4, 4) {$\mathcal{CJDF}$};
				\node at (1.5, 4.7) {$\mathcal{R}$};
				
			\end{tikzpicture}
			\label{fig:strong_def}
	}

\caption{Characteristics of deflagrations. For the weak deflagration, the characteristics (blue lines) impinge on the left side of the discontinuity only. To the right is a smooth transition zone (dashed red lines), on the left edge of which the reaction takes place.
In the case of an unstable strong deflagration, characteristics on either side of the discontinuity diverge. It is therefore replaced by a compound wave composed of a CJ deflagration (red lines) and a rarefaction (thick black lines and dashed blue lines). The dotted green line between them illustrates that the characteristics to the left of the CJ deflagration are parallel to the discontinuity -- in reality, the width of this region shrinks to zero and the deflagration and rarefaction waves are attached together.}
\label{fig:deflagration_characteristics}
\end{figure}
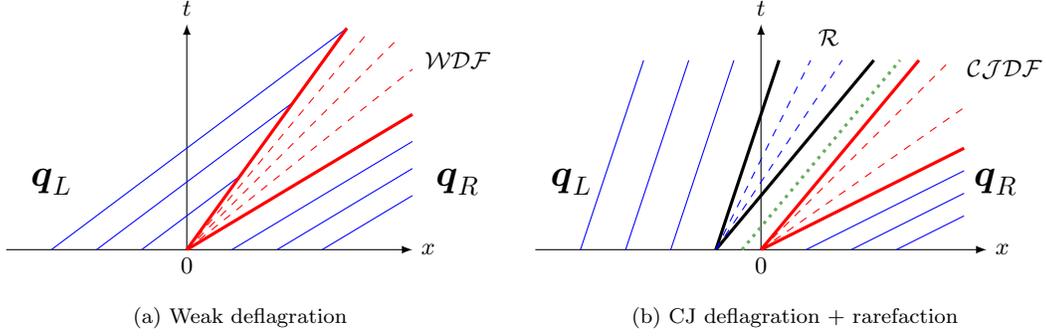

The compound wave starts with an inert precursor shock which raises the temperature of the material to the ignition temperature. This follows the exact equations \eqref{eq:exact_shock} of the shock, as in the detonation case, and we solve across the discontinuity using standard techniques.
Next, there is a deflagration wave, across which the reaction takes place and the pressure drops. Again, this follows the shock equations for a rarefaction, but with the same interpretation as in the detonation case.

The deflagration wave need not be stable. As with detonations, deflagrations fall into two classes: stable weak deflagrations and unstable strong deflagrations. For the stable weak deflagrations, the characteristic waves from one side impinge on the discontinuity, but not the other (see \fref{fig:weak_def}). For the unstable strong deflagrations, neither set of characteristic waves impinge on the discontinuity.

As for detonations, in the case of an unstable strong deflagration, the deflagration wave is replaced with a compound wave composed of a CJ deflagration (where the characteristics are parallel to the discontinuity) and a rarefaction wave (see \fref{fig:strong_def}).

In the calligraphic notation, we shall denote weak deflagrations by $\mathcal{WDF}$ and CJ deflagrations by $\mathcal{CJDF}$.

\subsection{Pressure-volume plot} \label{sec:newtonian_pressure_volume}
\begin{figure} \centering
\tikzsetnextfilename{RH_curves}
\begin{tikzpicture}[domain=0:10, scale=0.9]

	\draw[->] (-0.2,0) -- (10.2,0)  node[right] {$\hat{v}=\sfrac{1}{\hat{\rho}}$};
	\draw[->] (0,-0.2) -- (0,6.2) node[above] {$\hat{p}$};

	\draw[thick] plot[id=q, domain=0.3:10, smooth, samples=50] (\x, {(10 + 20 - \x) / (20*\x - 1.0)});
	\draw[dashed] (1,0) -- (1,1);
	\draw[dashed] (0,1) -- (1,1);
	\draw (1,0) node[below] {1};
	\draw (0,1) node[left] {1};

	\draw[thick, color=blue] plot[id=R1, domain=0:6.988] (\x, {-0.167*(\x-1.0) +1.0});
	\draw[thick, color=blue, dashed] plot[id=H1, domain=1:4.425] (\x, {-0.292*(\x-1.0) +1.0});
	\draw[thick, color=red] plot[id=R2, domain=0.18:1.167] (\x, {-6*(\x-1.0) +1.0});
	\draw[thick, color=red, dashed] plot[id=R2, domain=0:1] (\x, {-4.1876*(\x-1.0) +1.0});
	\draw[-] (1.6365, 0.8939) -- (2.6,1.8) node[right, font=\footnotesize] {weak deflagration};
	\fill[blue] (1.6365, 0.8939) circle(3pt);
	\draw[-] (5.7135, 0.2144) -- (6.5,0.5) node[right, font=\footnotesize] {strong deflagration};
	\fill[blue] (5.7135, 0.2144) circle(3pt);
	\draw[-] (2.3147, 0.6112) -- (3.5,1.1) node[right, font=\footnotesize] {CJ deflagration};
	\fill[blue] (2.3147, 0.6112) circle(3pt);
	\draw[-] (0.354, 4.876) -- (1.2, 5.1) node[right, font=\footnotesize] {strong detonation};
	\fill[red] (0.354, 4.876) circle(3pt);
	\draw[-] (0.871, 1.774) -- (1.8, 2.6) node[right, font=\footnotesize] {weak detonation};
	\fill[red] (0.871, 1.774) circle(3pt);
	\draw[-] (0.648, 2.454) -- (1.5, 3.4) node[right, font=\footnotesize] {CJ detonation};
	\fill[red] (0.648, 2.454) circle(3pt);

\end{tikzpicture}
\caption{Solutions of the reactive Riemann problem, as bounded by the Rayleigh lines (solid red and blue lines) and the Hugoniot curve (solid black curve). The point $(1,1)$, where $\hat{p}=\hat{v}=1$, corresponds to the known state. The dashed lines indicate the tangents to the Hugoniot curve passing through the known state, which intersect with the curve at the CJ points.}
\label{fig:RH_curves}
\end{figure}
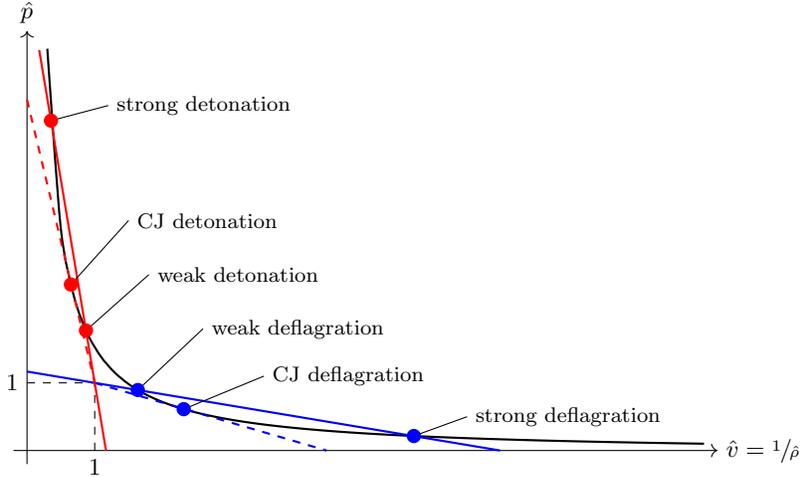

Another way to visualise the possible solutions for the reactive Riemann problem is to use a pressure-volume plot such as \fref{fig:RH_curves}, which follows \citep{Law2006}. Here, we plot the dimensionless pressure $\hat{p} = p_*/p_S$ against the specific volume $\hat{v} = \rho_S/\rho_*$, such that the known state $S$ corresponds to the point $(1,1)$. From the Rankine-Hugoniot relations, which relate the known state to the unknown state, we obtain equations for the Rayleigh lines (solid red and blue)
\begin{equation}\label{eq:newt_rayleigh}
\left(p_* - p_S\right)\left[\frac{1}{\rho_*}-\frac{1}{\rho_S}\right]^{-1} = -m^2,
\end{equation}
where the mass flux $m = \rho_Sv_S = \rho_*v_*$, and for the Hugoniot curve (solid black)
\begin{equation}\label{eq:newt_hugoniot}
\frac{\gamma}{\gamma-1}\left(\frac{p_*}{\rho_*} - \frac{p_S}{\rho_S}\right) - \frac{1}{2}\left(\frac{1}{\rho_*} + \frac{1}{\rho_S}\right)(p_* - p_S) = Q,
\end{equation}
where $Q$ is the reaction energy released per unit mass defined in \eqref{eq:energy_jump}.

The possible solutions are bounded by the intersections of the Rayleigh lines and the Hugoniot curve. The pressure jump across a detonation must therefore be somewhere between that of a weak detonation and a strong detonation (and similarly for deflagrations). The two tangent lines to the Hugoniot curve that pass through the known state intersect at the CJ points. These points separate strong detonations (deflagrations) from weak detonations (deflagrations), and occur where the flow in the unknown state is sonic, with velocity $v_* = c_{s,*}$.

\section{The relativistic reactive Riemann problem}\label{sec:rrrp}

In Newtonian hydrodynamics, the solution to the Riemann problem depends only on the normal component of any vector quantities in the initial conditions. However, in relativistic systems, the Lorentz factor introduces a coupling between the normal and tangential components. As found by \citet{Pons2000,Rezzolla2002} for the inert case, for high enough tangential velocities, the solution will smoothly transition from one wave pattern to another while maintaining the initial states otherwise unmodified. This is an entirely relativistic effect for which there is no Newtonian counterpart.

For the problem we are interested in (modelling neutron star oceans), we require general relativistic hydrodynamics. However, as described by \citet{Pons1998}, we need only consider the simpler \emph{special} relativistic Riemann problem. The Riemann problem for general relativistic fluids can be solved using special relativistic solvers following an appropriate coordinate transformation which transforms the coordinates at the interface to be locally Minkowskian. This follows from the equivalence principle, which states that physical laws in a local inertial frame for an arbitrary spacetime have the same form as those of special relativity. 

The introduction of this gauge transformation will change the component representation of the initial data and therefore change the detailed solution of the Riemann problem. The numerical values of, for example, the velocity components, should be interpreted as being in the local inertial frame, and therefore incorporate gravitational and rotational effects as seen by a distant observer. However, the qualitative wave pattern is gauge invariant. Therefore transforming to the local inertial frame and using special relativistic solvers will allow us to determine the qualitative wave pattern with gravity.  

In the following sections, we shall present the solutions for rarefactions and shocks, based on the equations given in \citet{Marti1994,Pons2000,Marti2003,Marti2015}.

\subsection{Rarefactions}
\label{sec:rel_raref}

Given the known state $S$ ahead of the rarefaction wave, we wish to calculate the unknown state behind the wave. 
Across a rarefaction, the normal velocity $v_x$ satisfies
\begin{equation}\label{eq:dvdp}
  \frac{\text{d} v_x}{\text{d} p} = \pm \frac{1}{\rho h W^2 c_s} \frac{1}{\sqrt{1 + g \left( \xi_{\pm}, v_x, v_t \right)}},
\end{equation}
where $v_t$ is the modulus of the tangential velocity, and we define the quantities
\begin{equation}
  g(\xi_\pm, v_x, v_t) = \frac{v_t^2 \left( \xi_{\pm} - 1 \right)}{\left( 1 - \xi_{\pm} v_x \right)^2},
\end{equation}
and
\begin{equation}
\xi_\pm = \frac{v_x(1-c_s^2) \pm c_s\sqrt{(1-v^2)[1-v^2c_s^2 - v_x^2(1-c_s)^2]}}{1-v^2c_s^2}.
\end{equation}
The $+(-)$ sign corresponds to $S=R$ $(S=L)$, and $c_s$ is the local sound speed. The local speed $v = \sqrt{v_x^2 + v_t^2}$ is used to calculate the Lorentz factor $W = (1-v^2)^{-1/2}$.

We also solve for the rest mass density and specific internal energy across the wave using
\begin{equation}
  \frac{\text{d}}{\text{d} p} \begin{pmatrix} \rho \\ \epsilon \end{pmatrix} = \frac{1}{h c_s^2}\begin{pmatrix} 1 \\ p/\rho^2 \end{pmatrix}.
\end{equation}
To find the unknown state, we connect the known state ahead of the wave to the unknown state behind the wave by integrating \eqref{eq:dvdp}. Using the fact that $hWv_t = \text{const}$ across a rarefaction wave, we can then calculate the tangential velocity as
\begin{equation}\label{eq:tangential_velocity}
  v_t = h_S W_S  v_{t,S} \left[\frac{1 - v_x^2}{h^2 + \left( h_S W_S v_{t,S} \right)^2}\right]^{\frac{1}{2}},
\end{equation}
where the $S$ subscript denotes the value of the variable in state $S$. The relation for the tangential velocity holds across both continuous and discontinuous waves.

\subsection{Shocks}
\label{sec:rel_shock}

Across a shock, the post-shock state can be found from the Taub adiabat \citep{Thorne1973}
\begin{equation}
\llbracket h^2\rrbracket = \left(\frac{h_b}{\rho_b} + \frac{h_a}{\rho_a}\right)\llbracket p\rrbracket,
\end{equation}
where $\llbracket q \rrbracket = q_b - q_a$. In general, this is solved using the equation of state to obtain the post-shock enthalpy as a function of the post-shock pressure, $h=h(p)$.

From this we can compute the mass flux $j$ across the shock
\begin{equation}
  j(p) = \sqrt{ \left(p_S - p\right)\left[ \frac{h_S^2 - h(p)^2}{p_S - p} - \frac{2 h_S}{\rho_S} \right]^{-1} }.
\end{equation}
This gives the shock velocities
\begin{equation}
  V_\pm(p) = \frac{ \rho_S^2 W_S^2 v_{x,S}  \pm j(p)^2 \sqrt{ 1 + \rho_S^2 W_S^2 \left( 1 - v_{x,S}^2 \right)/j(p)^2}}{\rho_S^2 W_S^2 + j(p)^2}.
\end{equation}

Given the shock velocity we can compute the shock Lorentz factor $W_S = (1 - V_\pm^2)^{-\sfrac{1}{2}}$, from which the post-shock normal velocity is
\begin{equation}
  v_x = \left( h_S W_S v_{x,S} \pm \frac{(p - p_S)}{j(p)\sqrt{1-V_\pm(p)^2}} \right) \left[ h_S W_S + \left( p - p_S \right) \left( \frac{1}{\rho_S W_S} \pm \frac{v_{x,S}}{j(p)\sqrt{1-V_\pm(p)^2}} \right) \right]^{-1}.
\end{equation}

The tangential velocity is the same as for the rarefaction wave in \eqref{eq:tangential_velocity}.

\subsection{Detonations \& deflagrations}
The solution procedure for reactive waves in the relativistic problem is the same as that for the Newtonian reactive Riemann problem discussed above in \sref{sec:reactive_rp}. The qualitative features are unchanged, with stable reactive waves taking the form of strong detonations, weak deflagrations or Chapman-Jouget detonations/deflagrations. Consequently, we can use the same solution procedure here as in the Newtonian case, first outlined by \citet{Zhang1989}.

In brief, if the pressure increases across a reactive wave we solve for a detonation. The relativistic shock relations of section~\ref{sec:rel_shock} are used, with the equation of state changing across the wave as described in the Newtonian case in section~\ref{sec:newt_detonation}. If the characteristics indicate that the detonation is unstable then a CJ detonation is found, and the relativistic rarefaction relations of section~\ref{sec:rel_raref} used to complete the compound wave.

If instead the pressure decreases across a reactive wave then we solve for a deflagration. The inert relativistic shock that matches the known state to the ignition temperature is found, and then the relativistic shock relations (with, as in the detonation case, the equation of state changing across the reactive wave) are used to find the deflagration. Again, if the characteristics indicate the deflagration is unstable then a CJ deflagration is found, and the compound wave solution completed with a relativistic rarefaction.

Note that, as we are solving in the lab frame in the relativistic case, it is easiest to identify the wave type from the characteristic structure directly. As shown in Figures~\ref{fig:detonation_characteristics} and \ref{fig:deflagration_characteristics}, the ordering of the wave speeds of the constant states either side of the deflagration or detonation directly fix the wave type. Detonations (or respectively deflagrations) can be identified by checking that the wave is supersonic (respectively subsonic), but this must be done by Lorentz boosting into the frame of the wave using, for example, the shock velocity as above.

\subsection{Pressure-volume plot}
As we did in \sref{sec:newtonian_pressure_volume} for the Newtonian case, we can visualise the solutions of the relativistic reactive Riemann problem using a pressure-volume plot \citep{Gao2012}. For relativistic flows, the Rayleigh and Hugoniot relations are given by 
\begin{eqnarray}\label{eq:relativistic_rayleigh}
(\hat{p} - 1) &=& -u_S^2\left[\hat{v} - 1 + \frac{\gamma}{\gamma -1}\left(\hat{p}\hat{v}^2 - 1\right) - \hat{q}\right],\\
\frac{\gamma + 1}{\gamma -1}(\hat{p}\hat{v} - 1) - (\hat{p} - \hat{v}) - 2\hat{q} &=& 
\frac{\gamma}{\gamma -1}\hat{p}(1-\hat{v}^2) 
- \frac{\gamma}{(\gamma-1)^2}(\hat{p}^2
\hat{v}^2 - 1) + \left(\hat{p} + \frac{\gamma +1}{\gamma -1}\right)\hat{q} + \hat{q^2},
\end{eqnarray}
where the normal 4-velocity for state \(S\) is \(u_S = v_{x,S}W_S\), \(\hat{p} = p_* / p_S\), \(\hat{v} = \rho_S / \rho_*\) and \(\hat{q} = \frac{\rho_S}{p_S}Q\). Comparing these to \eqref{eq:newt_rayleigh} and \eqref{eq:newt_hugoniot}, the key difference is the appearance of the \(\hat{q}\) term in the Rayleigh relation. This is because in the relativistic fluid equations, the reaction heat release term is present in the momentum conservation equation (through the enthalpy). A consequence of this is that the Rayleigh lines are now a function of \(\hat{q}\) and will no longer pass through the (1,1) point in the \(\hat{p}-\hat{v}\) diagram if \(\hat{q} \neq 0\). 

\begin{figure} \centering
\includegraphics[width=0.6\textwidth]{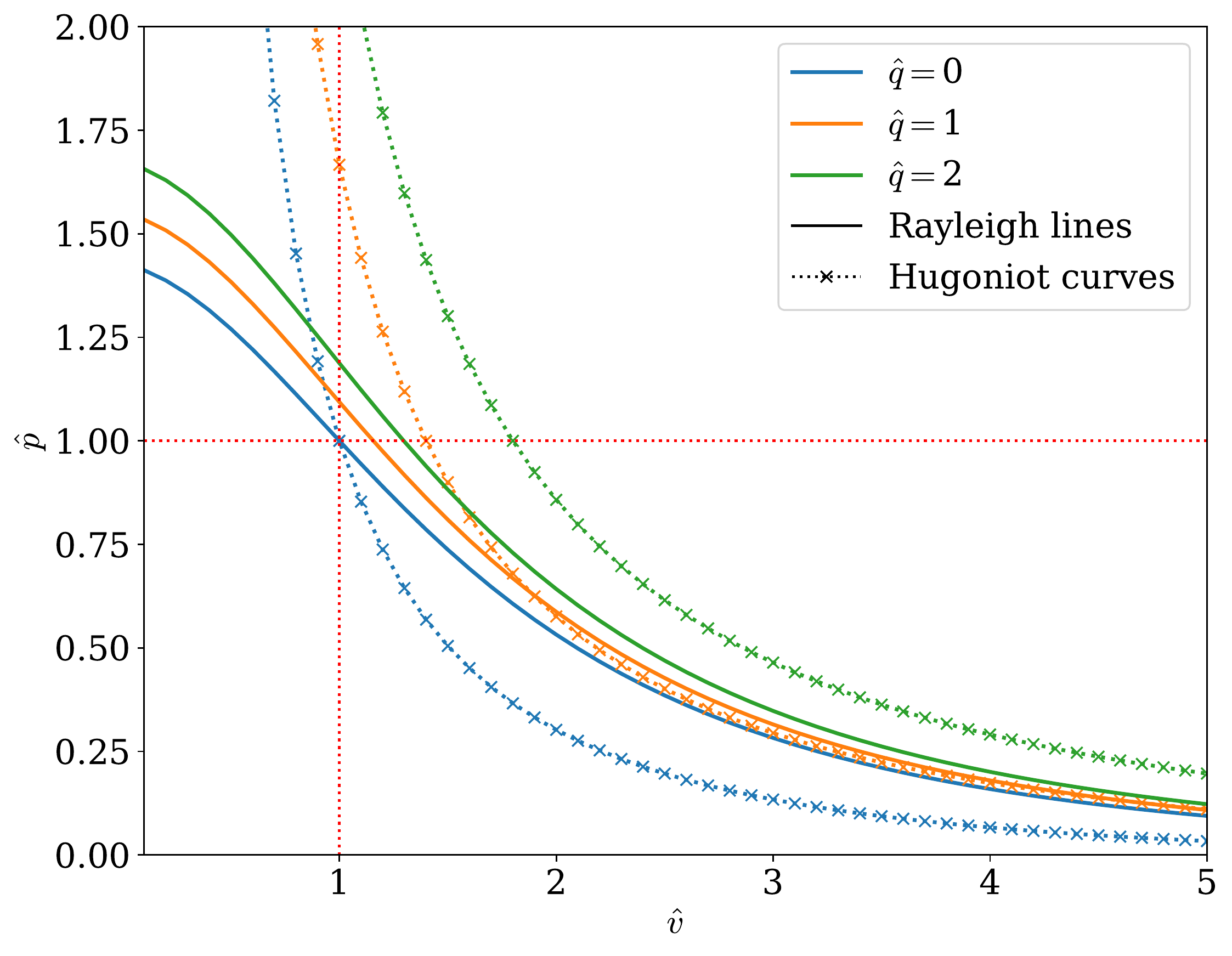}
\caption{Pressure-volume plot for the relativistic reactive Riemann problem for normal 4-velocity \(u_S = 0.35\), adiabatic index \(\gamma=\sfrac{5}{3}\) and various values of \(\hat{q}\). Plotted are the Rayleigh lines and the Hugoniot curves. Unlike for the Newtonian case, both the Rayleigh lines and the Hugoniot curves are functions of functions of \(\hat{q}\), with only the lines for \(\hat{q}\) intersecting the (1,1) point. As \(\hat{q}\) increases, both the Rayleigh lines and Hugoniot curves move rightwards, away from the (1,1) point.} 
\label{fig:pv_vary_q}
\end{figure}

This can be seen in \fref{fig:pv_vary_q}, where the Rayleigh lines and Hugoniot curves are plotted for various values of \(\hat{q}\). As \(\hat{q}\) increases, both the Rayleigh lines and Hugoniot curves move rightwards, away from the (1,1) point. By varying \(\hat{q}\), the possible solutions of the problem changes. For \(\hat{q} = 0\), there is a single intersection of the curves at (1,1): only weak deflagrations are a valid solution. For \(\hat{q} = 1\), there are two intersections, indicating that both weak and strong deflagrations are valid for the range of \(\hat{p}\) and \(\hat{v}\) where the Rayleigh line is above the Hugoniot curve. For \(\hat{q}=2\), there are no intersections. There are therefore no valid deflagration solutions for this initial data.

\begin{figure} \centering
\includegraphics[width=0.6\textwidth]{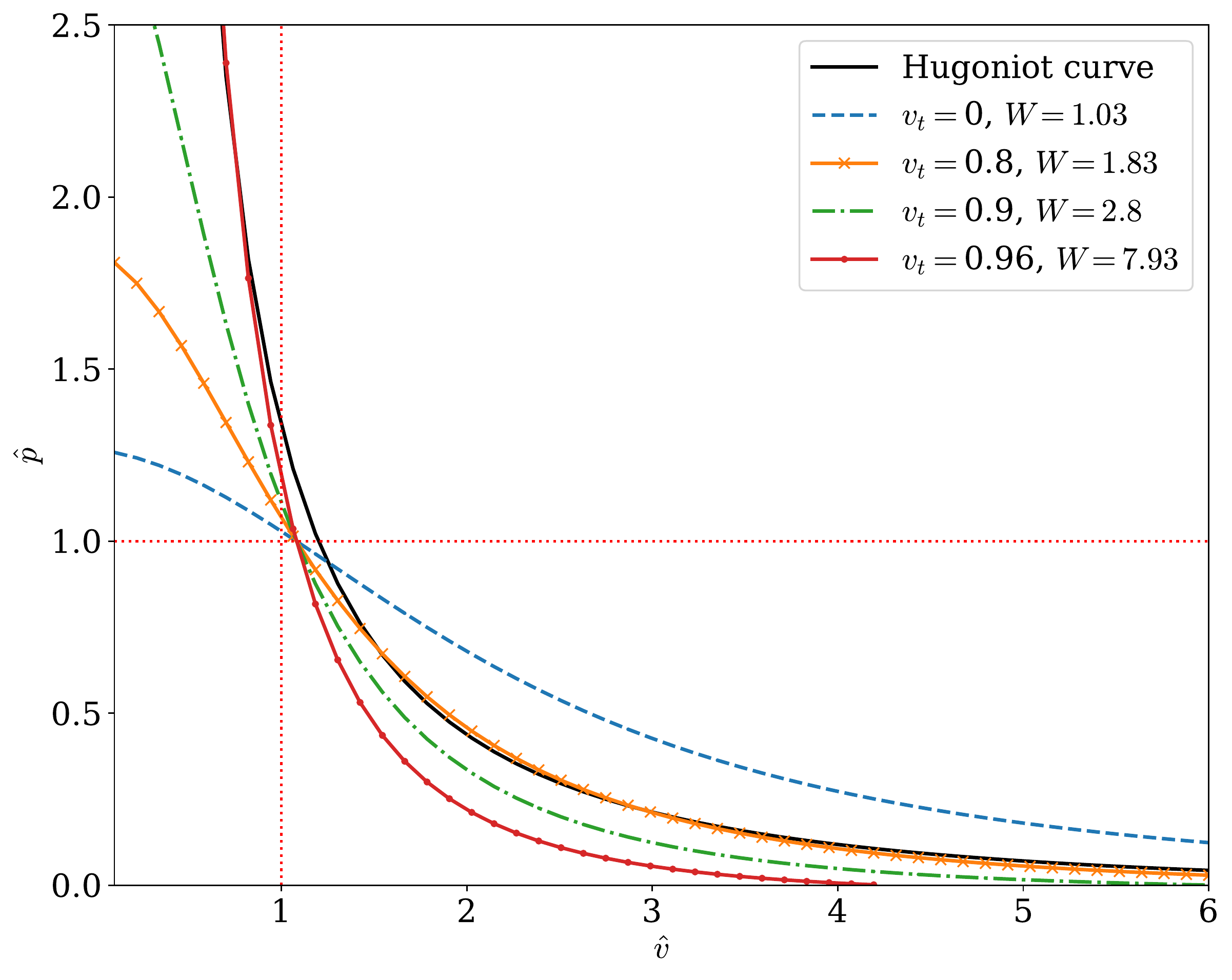}
\caption{Pressure-volume plot for the relativistic reactive Riemann problem for normal 3-velocity \(v_{x,S} = 0.25\) and \(\hat{q}=0.5\). Plotted are the Hugoniot curve and Rayleigh lines for various values of tangential velocity \(v_t\). Unlike for the Newtonian case, the coupling of the tangential velocity via the Lorentz factor means that the range of possible solutions changes with the tangential velocity.} 
\label{fig:pv_vary_vt}
\end{figure}

In the Newtonian Rayleigh relation \eqref{eq:newt_rayleigh}, the normal velocity appears in the mass flux term. In the corresponding relativistic relation \eqref{eq:relativistic_rayleigh}, the tangential velocity \(v_t\) appears due to the introduction of the Lorentz factor in the normal 4-velocity. This means that the Rayleigh curve is now a function of \(v_t\), and its intersection with the Hugoniot curve will change as \(v_t\) changes. This is illustrated in \fref{fig:pv_vary_vt}, where the Rayleigh lines are plotted for a range of tangential velocities. For \(v_t \lesssim 0.65\), there is only a single intersection of the Rayleigh line and the Hugoniot curve -- only weak deflagrations are possible for this system. For \(0.65 \lesssim v_t \lesssim 0.9\), the lines intersect twice, such that both weak and strong deflagrations are valid solutions for the range of \(\hat{p}\) and \(\hat{v}\) where the Rayleigh line is above the Hugoniot curve. For \(v_t \gtrsim 0.9\), there are no intersections of the curves in the bottom right deflagration quadrant, and so no deflagration solutions exist for the problem. However, it can be seen for the curve with \(v_t = 0.96\) that if the tangential velocity is increased still further, the lines will now intersect in the detonation quadrant of the diagram. Weak detonations are therefore now valid solutions of the problem. As the Lorentz factor \(W\rightarrow \infty\), the Rayleigh and Hugoniot curves still only intersect once. There are therefore no possible strong detonation solutions for this system for any value of the tangential velocity. 

\section{Numerical results}\label{sec:numerics}

In order to investigate the reactive relativistic Riemann problem, we developed the numerical solver, \texttt{R3D2}. This open source Python-based code solves the equations outlined in the sections above for the inert and reactive relativistic Riemann problem. A detailed description of the code and instructions for its usage can be found in \citet{Harpole2016}.

The following section contains results of simulations performed for a range of 1d systems using \texttt{R3D2} in order to demonstrate the features of the inert and reactive relativistic Riemann problem. We shall show how the solution changes with tangential velocity and reaction terms. 

In this section, we shall restrict ourselves to the gamma law equation of state,
\begin{equation}
p = (\gamma -1)\rho \varepsilon,
\end{equation}
and assume that the heat release parameter \(q\) is constant. These are reasonable assumptions to make when the reactions take place over extremely short timescales, as we assume they do in the reactive Riemann problem. In other words, we are effectively treating the real problem using constant \(\gamma\) and \(q\) on local scales where the effects of e.g.~turbulence, shock curvature and more complex burning can be neglected.

The results presented in this section are in `code units', where we set the speed of light $c=1$. As a result, lengths and times have the same dimension, as do mass and energy densities: $[\rho] = [p] = [q]$ \citep{Marti2015}.

\subsection{Inert relativistic Riemann problem}

\begin{figure*} \centering
\includegraphics[width=0.8\textwidth]{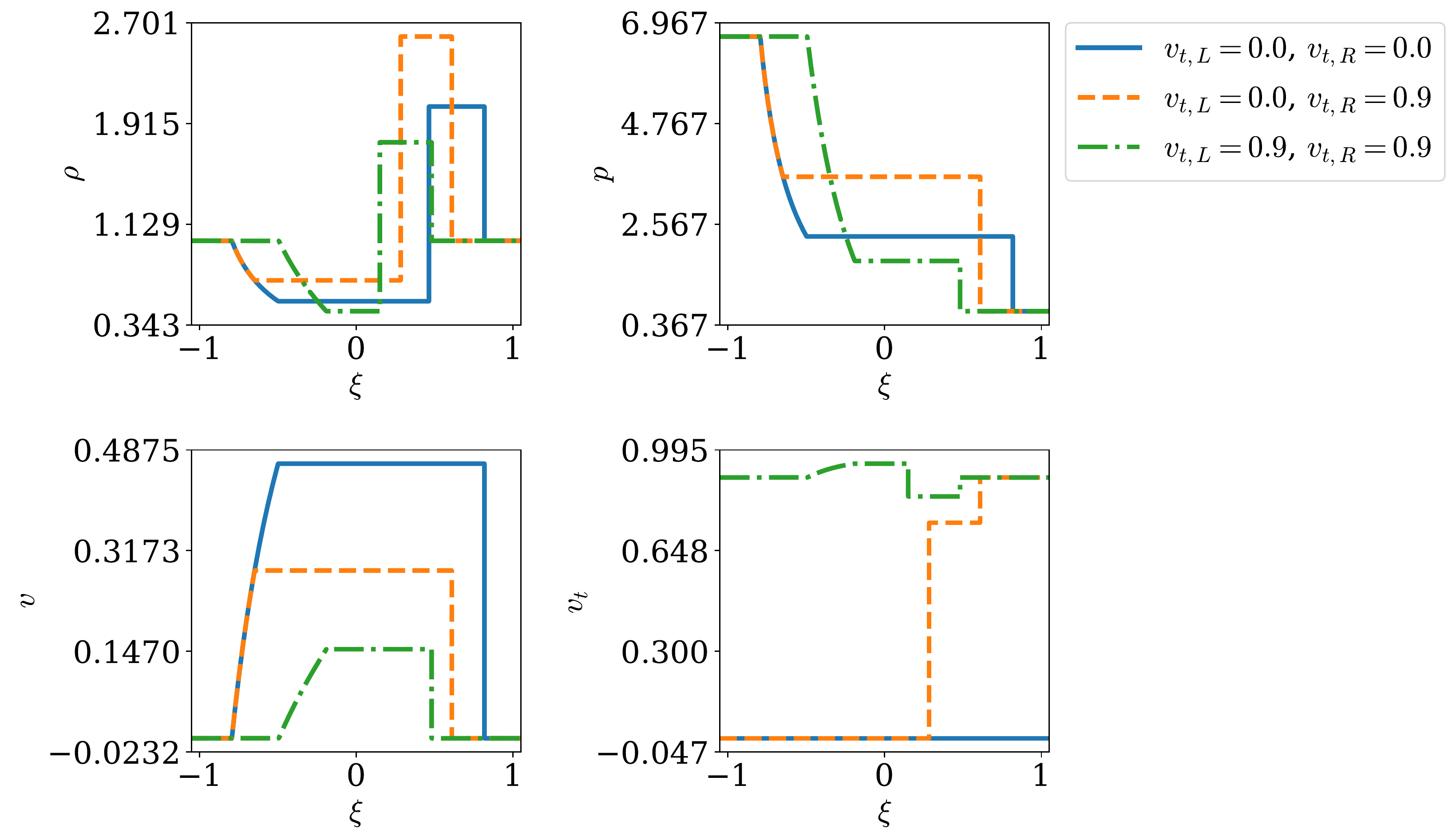}
    \caption{Exact solution of the inert relativistic Riemann problem for different initial tangential velocities, calculated using \texttt{R3D2}. The variables are plotted as a function of the dimensionless, scale-free parameter $\xi = x / t$, as introduced in \eqref{eq:xi}. All cases share the initial conditions $(\rho, v_x, \varepsilon)_L = (1, 0, 10)$, $(\rho, v_x, \varepsilon)_R = (1, 0, 1)$. For the blue curves, $v_t (L, R) = (0,0)$, for the orange dashed curves, $v_t (L, R) = (0,0.9)$, and for the green dot-dashed curves, $v_t (L, R) = (0.9,0.9)$. The final state consists of a left-going rarefaction wave, a contact wave and a right-going shock wave: $\mathcal{R}_\leftarrow \,\mathcal{C}\, \mathcal{S}_\rightarrow$. Increasing the tangential velocity alters the intermediate state, reducing the normal velocity.}
\label{fig:varying_vt}
\end{figure*}

The effect of tangential velocity in the inert relativistic Riemann problem can be seen in \fref{fig:varying_vt}. In this example, the final state is made up of a left-going rarefaction wave, a contact wave and a right-going shock wave: $\mathcal{R}_\leftarrow \,\mathcal{C}\, \mathcal{S}_\rightarrow$. We can see that the introduction of a tangential velocity reduces the width of the intermediate state and changes the intermediate wave states. In the case with a non-zero right state tangential velocity $v_t (L, R) = (0,0.9)$, the density and pressure in the intermediate states is increased, whereas in the case with non-zero tangential velocities for both states $v_t (L, R) = (0.9,0.9)$, they are decreased. For both cases, the normal velocity in the intermediate states is reduced. As stated above and first noted by~\citet{Pons2000,Rezzolla2002}, this effect is a purely relativistic one that is not present in the Newtonian Riemann problem, where the tangential velocity has no effect on the qualitative wave pattern in the solution. 

\subsection{Reactive relativistic Riemann problem}

\begin{figure*} \centering
\includegraphics[width=0.8\textwidth]{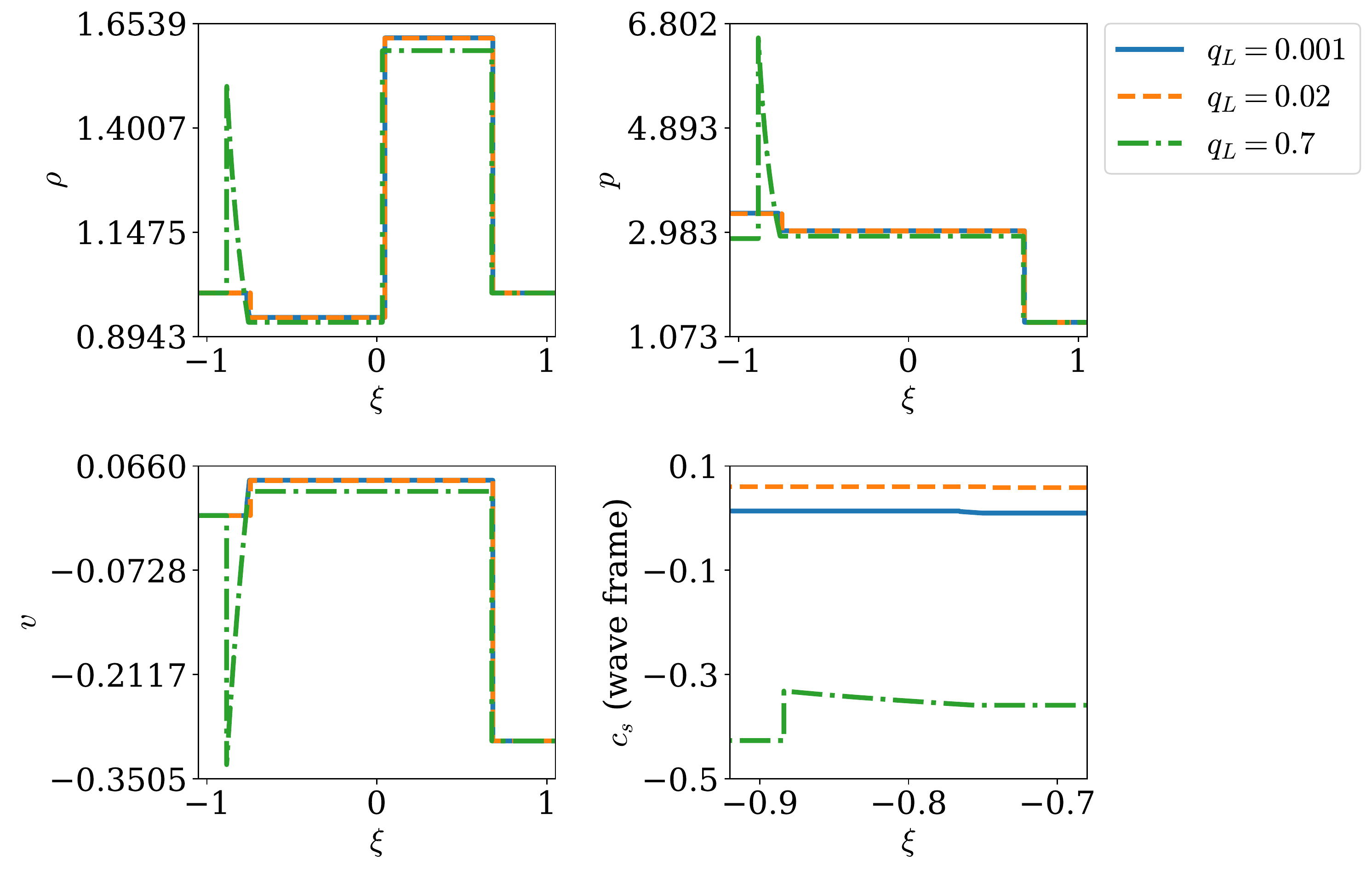}
\caption{Deflagration to detonation transition for different values of \(q = (0.001, 0.02, 0.7)\), corresponding to the solid blue, dashed orange and dotted green curves respectively. The initial conditions are \((\rho, v_x, v_t, \varepsilon)_L = (1, 0, 0, 5.0)\), \((\rho, v_x, v_t, \varepsilon)_R = (1, -0.3, 0, 2)\). For \(q=0.001\), the solution is a CJ deflagration, \(\mathcal{(CJDF_\leftarrow\,R_\leftarrow)C\,S_\rightarrow}\). As \(q\) is increased to \(q=0.02\), the CJ deflagration becomes a weak deflagration, \(\mathcal{WDF_\leftarrow\, C\,S_\rightarrow}\), and at \(q=0.7\) this has transitioned to a CJ detonation, \(\mathcal{(CJDT_\leftarrow\,R_\leftarrow)C\,S_\rightarrow}\). Note that the plot of the sound speed (Lorentz boosted into the frame of the wave) in the lower right has different $x$-axis limits to the other plots: we have zoomed in about the reactive wave to highlight the behaviour in this region. }
\label{fig:varying_q_ddt}
\end{figure*}

The wave pattern of the solution to the reactive Riemann problem can be sensitive to initial conditions. In \fref{fig:varying_q_ddt}, the effect on a detonation of changing the energy of reaction $q$ can be seen. In this example, when \(q=0.001\), the solution contains a left-going CJ deflagration \(\mathcal{(CJDF_\leftarrow\,R_\leftarrow)C\,S_\rightarrow}\). Increasing the reaction energy causes the left-going reactive wave in the system to transition to a weak deflagration when \(q=0.02\), \(\mathcal{WDF_\leftarrow\, C\,S_\rightarrow}\), then on to a CJ detonation when \(q=0.7\), \(\mathcal{(CJDT_\leftarrow\,R_\leftarrow)C\,S_\rightarrow}\). This transition can be seen by considering the lower right plot of the soundspeed, Lorentz boosted into the frame of the wave (i.e., the speed of the detonation, or the largest speed associated with a compound wave). Across the deflagrations, moving from the burnt fluid on the right to the unburnt fluid on the left, the sound speed increases. Across the detonation, the sound speed instead decreases. For detonations the wave is travelling left faster than the sound speed -- the wave is supersonic -- whilst for deflagrations it is subsonic.

\subsection{Tangential velocity}\label{sec:tangential_velocity}

As described above, \citet{Pons2000,Rezzolla2002} found that sufficiently high tangential velocities can change the wave pattern in the inert relativistic Riemann problem. Here we investigate this for the reactive relativistic system to see whether it could be relevant for X-ray bursts on rapidly rotating neutron stars: given the conditions in the ocean, would it be possible for a fast enough tangential velocity to develop to induce a transition from deflagration to detonation? As discussed in \sref{sec:rrrp}, although we are interested in general relativistic effects, by using a suitable coordinate transformation we can investigate these by considering the special relativistic problem. 

\begin{figure*} \centering
\includegraphics[width=\textwidth]{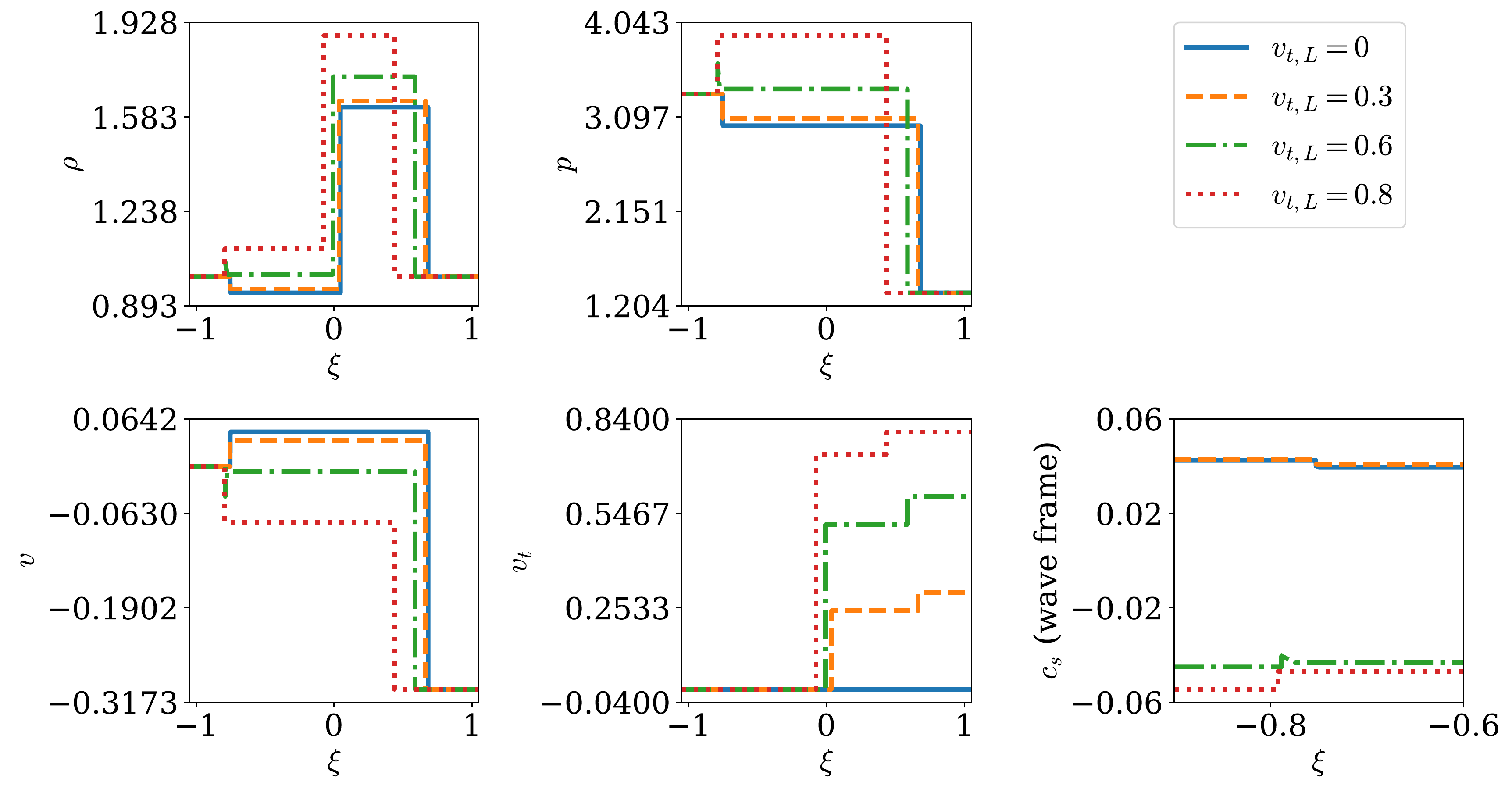}
\caption{Deflagration to detonation transition for different values of \(v_{t,R} = (0.0, 0.3, 0.6, 0.8)\), corresponding to the solid blue, dashed orange, dash-dotted green and dotted red lines respectively. The initial conditions are \((\rho, v_x, v_t, \varepsilon)_L = (1, 0, 0, 5.0)\), \((\rho, v_x, v_t, \varepsilon)_R = (1, -0.3, v_t, 2)\). For \(v_{t,R} = 0\), the solution is \(\mathcal{(CJDF_\leftarrow\,R_\leftarrow)\,C\,S_\rightarrow}\). At \(v_{t,R} = 0.3\), the CJ deflagration has transitioned to a weak deflagration, producing the solution \(\mathcal{WDF_\leftarrow\,C\,S_\rightarrow}\). At \(v_{t, R} = 0.6\), the deflagration has transitioned to a CJ detonation, \(\mathcal{(CJDT_\leftarrow\,R_\leftarrow)\,C\,S_\rightarrow}\), then at \(v_{t, R} = 0.8\), this has transitioned to a strong detonation, \(\mathcal{SDT_\leftarrow\,C\,S_\rightarrow}\). As seen in the previous plot, across the deflagrations the soundspeed increases moving leftwards from the burnt to the unburnt fluid, whereas for the detonations it decreases, and in the frame of the wave we see the detonations are supersonic and the deflagrations subsonic. }
\label{fig:varying_vt_ddt}
\end{figure*}

As seen in \fref{fig:varying_vt_ddt}, we found
that varying the tangential velocity can indeed change the wave pattern, causing the reactive wave to transition from a deflagration to a detonation. In the particular system shown, the reactive wave takes all possible forms from CJ deflagration to strong detonation as the tangential velocity is increased. This is not the case for all systems. Clearly, for a system already containing a strong detonation when \(v_t=0\), increasing the tangential velocity will only make the detonation stronger. However, even for systems containing deflagrations when \(v_t=0\), increasing the tangential velocity may not be sufficient to induce a transition -- the system already has to be sufficiently close to transitioning for the increase in tangential velocity to be able to `tip it over the edge'. 

In order to determine whether or not this effect may be relevant in neutron star oceans, we must consider initial data that represents the properties of the system. To do this, we must convert physical quantities into our `code units'. As described by \citet{Marti2015}, this is done by completing the system of units with two independent reference quantities (in addition to the velocity unit, $c=1$). Here we shall use a reference density and temperature, setting \(\rho_r = \SI{e5}{\gram\per\cubic\centi\metre} = 1\) and \(T_r = \SI{e9}{\kelvin} = 1\). Using thermodynamic quantities typical of the neutron star ocean at the depth where helium burning would occur, we get the rescaled values listed in~\tref{tab:ns_ocean_properties}. In the case of a photospheric radius expansion (PRE) burst, the height of the ocean increases by a factor of \(\sim 10\). This corresponds to a decrease in the density of the burnt material by a factor of 10. 

\begin{table}
\centering
\begin{tabular}{>{$}c<{$} c c}
\hline
\hline
\text{Quantity} & \text{cgs} & \text{rescaled} \\
\hline
\rho & \SI{e5}{\gram\per\cubic\centi\metre} & \(\rho_r\)\\
p & \SI{4e22}{\erg\per\cubic\centi\metre} & \SI{4e-4}{\rho_r.c^2} \\
q & \SI{6e17}{\erg\per\gram} & \SI{6e-4}{c^2}\\
\varepsilon & \SI{8e17}{\erg\per\gram} & \SI{8e-4}{c^2}\\
T & \SI{e9}{\kelvin} & \(T_r\)\\
\hline
\end{tabular}
\caption{Typical properties of the neutron star ocean at a depth where helium burning would occur \protect\citep{Strohmayer2001,Cavecchi2013a} in cgs units and rescaled in units, where we set \(c=\SI{3e10}{\centi\metre\per\second} = 1\), \(\rho_r = \SI{e5}{\gram\per\cubic\centi\metre} = 1\) and \(T_r = \SI{e9}{\kelvin} = 1\). }
\label{tab:ns_ocean_properties}
\end{table}

\begin{figure*} \centering
\includegraphics[width=0.8\textwidth]{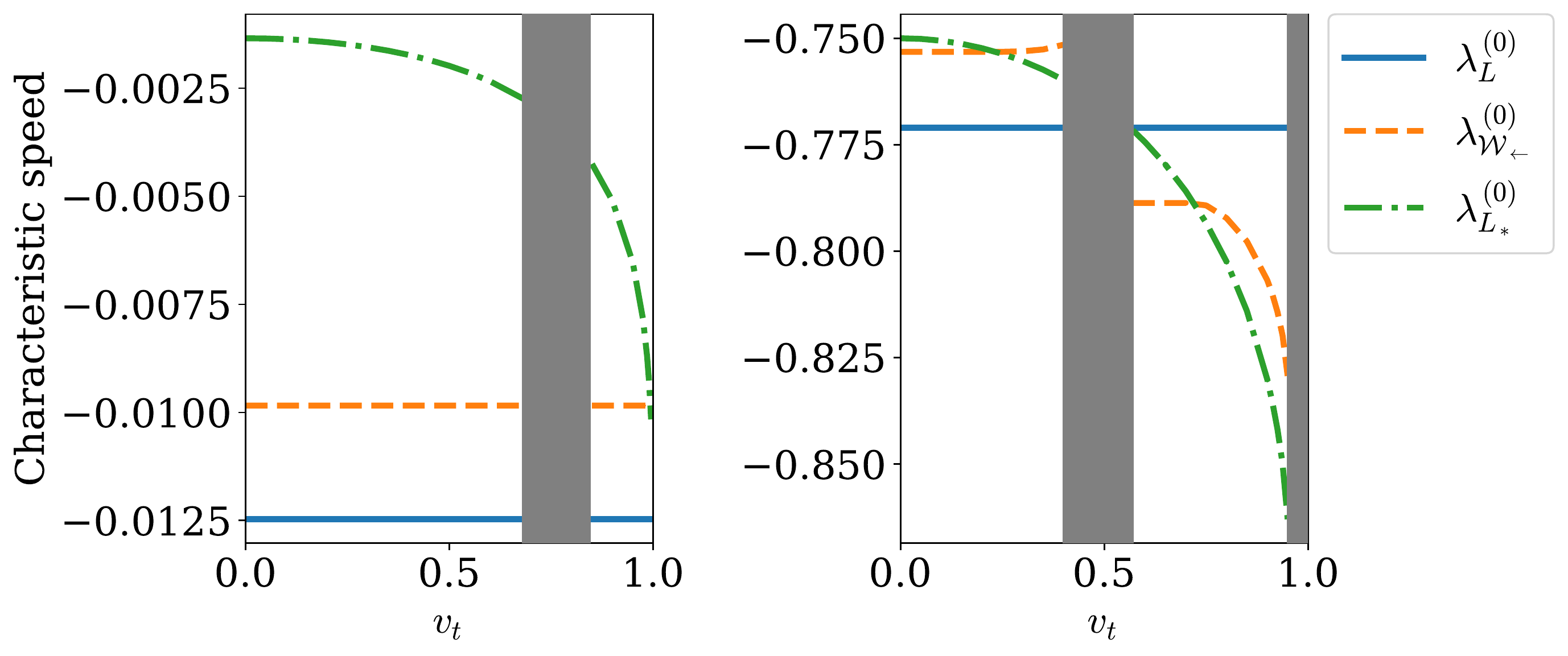}
\caption{As can be seen from Figures~\ref{fig:detonation_characteristics} and \ref{fig:deflagration_characteristics}, deflagration to detonation transitions occur when the characteristic wave structure changes. The left plot shows the smallest characteristic speeds ($\lambda^{(0)}$) of the states either side of the reactive wave ($\lambda^{(0)}_{L}, \lambda^{(0)}_{L_*}$), and the speed of the left-going reactive wave ($\lambda^{(0)}_{{\cal W}_\leftarrow}$) for the `realistic' initial data $\left(\rho, \varepsilon, q, C_v, T_i\right)_L = \left(1, 1.5\times 10^{-4}, 10^{-5}, 10^{-4}, 1\right)$, $\left(\rho, \varepsilon\right)_R = \left(0.1, 1.5\times 10^{-4}\right)$. The right plot shows the characteristic speeds for the initial data used in \fref{fig:varying_vt_ddt}. The grey shaded regions are where no physical solution can be found, either because the characteristic structure is unphysical ($0.4 \lesssim v_t \lesssim 0.6$) or the total velocity of one state would be superluminal ($0.95 \lesssim v_t$). For small $v_t$ both plots show a CJ deflagration: the speeds are ordered $\lambda^{(0)}_{L} < \lambda^{(0)}_{{\cal W}_\leftarrow} < \lambda^{(0)}_{L_*}$ to match panel (b) of Figure~\ref{fig:deflagration_characteristics}. Both plots also show a transition to a weak deflagration (at $v_t \sim 0.25$ in the right plot, and at $v_t \sim 1$ in the left) when $\lambda^{(0)}_{L} < \lambda^{(0)}_{L_*} < \lambda^{(0)}_{{\cal W}_\leftarrow}$, to match the (mirror image of) panel (a) of Figure~\ref{fig:deflagration_characteristics}. Only the right plot shows the wave structure changing to a strong detonation (at $v_t\sim 0.7$) when $\lambda^{(0)}_{L_*} < \lambda^{(0)}_{{\cal W}_\leftarrow} < \lambda^{(0)}_{L}$, matching panel (a) of Figure~\ref{fig:detonation_characteristics}. No such deflagration to detonation transition appears in the left plot for the initial data consistent with `realistic' X-ray bursts.}
\label{fig:char_speed_vs_ddt}
\end{figure*}

We can use this rescaled data to model the ocean waves near a burning front associated with a `realistic' X-ray burst using the initial left and right unburnt and burnt states
\begin{eqnarray}
\left(\rho, \varepsilon, q, C_v, T_i\right)_L &=& \left(1, 1.5\times 10^{-4}, 10^{-5}, 10^{-4}, 1\right),\\
\left(\rho, \varepsilon\right)_R &=& \left(0.1, 1.5\times 10^{-4}\right),
\end{eqnarray}
where \(C_v\) is the heat capacity at constant volume and \(T_i\) is the ignition temperature. We assume the system to initially be static, with \(v_x = v_t = 0\).
This results in a CJ deflagration propagating at \(\sim \SI{e-3}{c}\) leftwards, which is consistent with a burst rise time of \(\sim \SI{1}{\second}\) and with predictions by \citet{Spitkovsky2002} for the maximum flame speed. Increasing the tangential velocity until it approaches the speed of light, the deflagration remains a CJ deflagration, but it is not possible to produce a transition to a detonation by changing the tangential velocity alone. This can be seen graphically in \fref{fig:char_speed_vs_ddt}, which shows how the order of the characteristic speeds neighbouring the reactive wave, and the wave speed of the reactive wave itself, remain unchanged as the tangential velocity approaches the speeds of light. From this we can conclude that conditions in the neutron star ocean are simply too far away from the conditions required for a detonation for a high tangential velocity to be able to induce a deflagration to detonation transition.

\section{Summary}

In neutron star oceans, it is conceivable that a burning front moving latitudinally may encounter a region with a fast tangential velocity produced by the star's rotation. In \sref{sec:tangential_velocity}, we investigated whether a fast tangential velocity may be able to cause a deflagration wave to transition to a detonation. To do this, we considered the relativistic reactive Riemann problem. Unlike for the Newtonian Riemann problem where only the velocity perpendicular to the interface is relevant, in the relativistic case the tangential velocity can become significant through the Lorentz factor. It was found that such a transition is possible, but for systems already on the verge of transitioning. Consequently, it is unlikely that such a transition would occur for a burning front in a neutron star ocean through this mechanism alone. 

In this work we have only considered the 1d problem, focusing on the specific relativistic effects that could lead to a transition to detonation. However, the expectation is that burning in neutron star oceans is highly turbulent, leading to highly wrinkled, multidimensional flame fronts. We have assumed the flame front to be an infinitesimally thin discontinuity, whereas such a turbulent flame would be significantly broadened due to the disruption of the reaction zone by turbulent eddies \citep{Peters2001}. Above, we modelled the burning as an instantaneous one-step reaction, whereas the real system involves complex microphysics with multiple species \citep{Wallace1981a}. The transverse velocity may produce cellular structures which can then interact, leading to a transition to detonation \citep{Clavin2012,Han2017}. In order to fully rule out the possibility of a deflagration to detonation transition in this system, we therefore need to carry out more realistic multidimensional simulations which take these effects into account. This investigation has been begun by the lead author in \citet{Harpole2018a}, and we plan to continue with this in the future.

Relativistic deflagrations also occur in other astrophysical systems such as supernovae and gamma ray bursts. However, given the combination of extreme conditions required for the transition to take place, it seems unlikely that this transition would occur in these systems either. As we have seen, not only does there need to be some process which generates an extremely high, relativistic tangential velocity (preferably over an extended period of time in order to maximise the chances of the reaction occurring), but the fluid must also be very hot and on the verge of reacting (without already having done so). The fluid must also be dense enough that burning in the fluid with zero tangential velocity propagates as a deflagration rather than a detonation. 

Core-collapse supernovae involve explosive burning of fast moving material. The progenitor stars can have very high rotation rates, so it may be possible for a fast tangential velocity to develop. However, even taking an optimistic estimate of the surface rotational velocity of the progenitor of $\sim \SI{300}{\kilo\metre\per\second}$, this is still only a fraction of a percent of the speed of light: any relativistic effects produced by this tangential velocity would therefore be negligible. The burning is also believed to propagate as a detonation from the start, so there would be no opportunity for it to transition.

Black hole accretion disks are another system where hot, burning material moving at high speeds can be found. Instabilities in the inner accretion disk are believed to be a source of gamma ray bursts \citep{Perna2006}. If the accretion disk was rotating fast enough, and unreacted material from the black hole jet were to fall back onto the disk, the necessary conditions could potentially be reached for the transition to occur. We can get an estimate for the maximum tangential velocity by considering the tangential component of the Keplerian velocity at the Schwarzschild radius. For a black hole of mass $M$, this is given by $v_t(R=R_S) = \frac{1}{2}c^2/\sqrt{GM}$. As this is inversely proportional the black hole's mass, we can find an upper bound on this by considering a small, solar mass black hole. This gives a tangential velocity of $v_t \sim 0.01c$, so even for this extreme example, the disk's tangential velocity will only be 1\% of the speed of light. Tangential flows in black hole accretion disks are therefore unlikely to be able to reach sufficiently high velocities for relativistic tangential effects to become important.

\acknowledgments

A.~Harpole thanks the UK Science and Technology Facilities Council (STFC) grant~1522890 for supporting her PhD. A Jupyter notebook containing the code to generate the results in this work can be found at \url{https://github.com/harpolea/r3d2/tree/master/examples}.

%

\vspace{5mm}


\software{r3d2 \citep{Harpole2016}, 
          numpy \citep{Oliphant2006}, 
          scipy \citep{Jones}, 
          matplotlib \citep{Hunter2007},
          Jupyter \citep{Kluyver2016}
          }

\bibliography{main}



\end{document}